*“This document is the Accepted Manuscript version of a Published Article that appeared in final form in Boopathi, K. M.; Martín-García, B.; Ray, A.; Pina, J. M.; Marras, S.; Saidaminov, M. I.; Bonaccorso, F.; Di Stasio, F.; Sargent, E. H.; Manna, L.; Abdelhady, A. L. Permanent Lattice Compression of Lead-Halide Perovskite for Persistently Enhanced Optoelectronic Properties. ACS Energy Letters 2020, 5 (2), 642–649, Copyright © 2020 American Chemical Society. To access the final published article, see: https://doi.org/10.1021/acsenergylett.9b02810.”*

# Permanent Lattice Compression of Lead-Halide Perovskite for Persistently Enhanced Optoelectronic Properties

Karunakara Moorthy Boopathi,[1] Beatriz Martín-García,[2] Aniruddha Ray,[1,3] Joao M. Pina,[4] Sergio Marras,[5] Makhsud I. Saidaminov, [4,6] Francesco Bonaccorso,[2,7] Francesco Di Stasio,[1] Edward H. Sargent,[4] Liberato Manna, [1]* Ahmed L. Abdelhady[1]*

[1]Nanochemistry Department, Istituto Italiano di Tecnologia, Via Morego 30, Genova 16163, Italy. Email: liberato.manna@iit.it, ahmed.abdelhady@iit.it

[2]Graphene Labs, Istituto Italiano di Tecnologia, Via Morego 30, 16163 Genova, Italy

[3]Dipartimento di Chimica e Chimica Industriale, Università degli Studi di Genova, Via Dodecaneso, 31, Genova 16146, Italy

[4]Department of Electrical and Computer Engineering, University of Toronto, Toronto, Ontario M5S 1A4, Canada

[5]Materials Characterization Facility, Istituto Italiano di Tecnologia, Via Morego 30, Genova 16163, Italy

[6]Department of Chemistry and Electrical & Computer Engineering, Centre for Advanced Materials and Related Technologies (CAMTEC), University of Victoria, 3800 Finnerty Road, Victoria, British Columbia V8P 5C2, Canada

[7]BeDimensional Srl., Via Albisola 121, 16163 Genova, Italy

***Abstract**. Under mild mechanical pressure, halide perovskites show enhanced optoelectronic properties. However, these improvements are reversible upon decompression, and permanent enhancements have yet to be realized. Here, we report antisolvent-assisted solvent acidolysis crystallization that enables us to prepare methylammonium lead bromide single crystals showing intense emission at all four edges under ultraviolet light excitation. We study structural variations (edge-vs-center) in these crystals using micro-X-ray diffraction and find that the enhanced emission at the edges correlates with lattice compression compared to in the central areas. Time-resolved photoluminescence measurements show much longer-lived photogenerated carriers at the compressed edges, with radiative component lifetimes of ∼1.4 µs, 10 times longer than at the central regions. The properties of the edges are exploited to fabricate planar photodetectors exhibiting detectivities of $3 \times 10^{13}$ Jones, compared to $5 \times 10^{12}$ Jones at the central regions. The enhanced lifetimes and detectivities correlate to the reduced trap state densities and the formation of shallower traps at the edges due to lattice compression.*

Over the past few years, hybrid organic–inorganic lead halide perovskites ($APbX_3$; A = alkylammonium and X = halide) have been investigated for solar cells, (1−7) light emitting diodes, (8−12) lasers, (13−15) and photodetectors. (16−22) Compared to polycrystalline thin films, perovskite single crystals show lower trap state densities and longer carrier lifetimes. (17,23−28) However, both perovskite thin films and single crystals suffer from relatively low photoluminescence (PL) due to the presence of sub-bandgap traps. (29−33)

Several reports describe $MAPbX_3$ (MA = methylammonium) perovskites as nonstoichiometric; they are halide-deficient. (34−37) The resultant defects, arising from halide vacancies and the presence of metallic lead, act as nonradiative recombination centers. (38,39) It has also been demonstrated that lattice strain can lead to nonradiative recombination. (40) The use of mixed cations/halides to induce strain relaxation in halide perovskites was found to reduce nonradiative recombination. (41−44) The incorporation of a foreign cation into the lattice is a means of inducing chemical pressure, one leading to lattice compression. (45,46) Under mild mechanical pressure, lattice compression of the halide perovskites leads to bandgap narrowing, increased PL intensity, a longer carrier lifetime, and shallower trap states, (47−57) but these effects disappear upon decompression. Hence, developing permanent compression is desired to maintain the enhanced optoelectronic properties.

Here, we report an antisolvent-assisted solvent acidolysis crystallization (AA-SAC) method to grow high quality, single metal-ion and single organic cation, $MAPbBr_3$ perovskite crystals at room temperature. These crystals evidenced intense emission at all four edges under ultraviolet (UV) excitation. Structural, compositional, and optical properties of the perovskite single crystals confirmed that the intense emission and the long PL lifetimes at the edges are due to compression with respect to the central regions of the crystals. Furthermore, the edges of the $MAPbBr_3$ crystals, when utilized in planar photodetectors, displayed excellent light detectivities, indicating low trap state densities.

We have developed a room temperature antisolvent assisted solvent acidolysis crystallization (AA-SAC) method to prepare high-quality perovskite single crystals, as shown in Figure 1a. Here, *N*-methylformamide (NMF) was used as both solvent and a source of MA cations through the hydrolysis of amides under acidic conditions, (58−60) while toluene was used as the antisolvent. The crystals grew mostly in the lateral dimensions and eventually became a millimeter-sized rectangular or square sheet with a thickness of 250 ± 20 µm (Figure S2). We monitored the crystal growth by taking pictures at different time intervals once the crystal was observed. As is shown in Figure S3, initially, the crystal was only a few microns large. The crystal growth was slightly faster in one direction, and at 100 min, it reached roughly $2 \times 1.5\ mm^2$. After an extra 140 min inside the growth solution the crystal became $3 \times 2\ mm^2$ in size. We show that the acidolysis of NMF by HBr is a gradual process. By using an antisolvent (tetrahydrofuran in this case) to precipitate out the perovskite, we noticed that the first precipitation takes place only 2 h after dissolving $PbBr_2$ in the NMF/HBr mixture. Afterward, the mass of the precipitated perovskite powder increased over time. The concentration of the *in situ* formed MA ions over time was calculated from the mass of the precipitated powders (Figure S4).

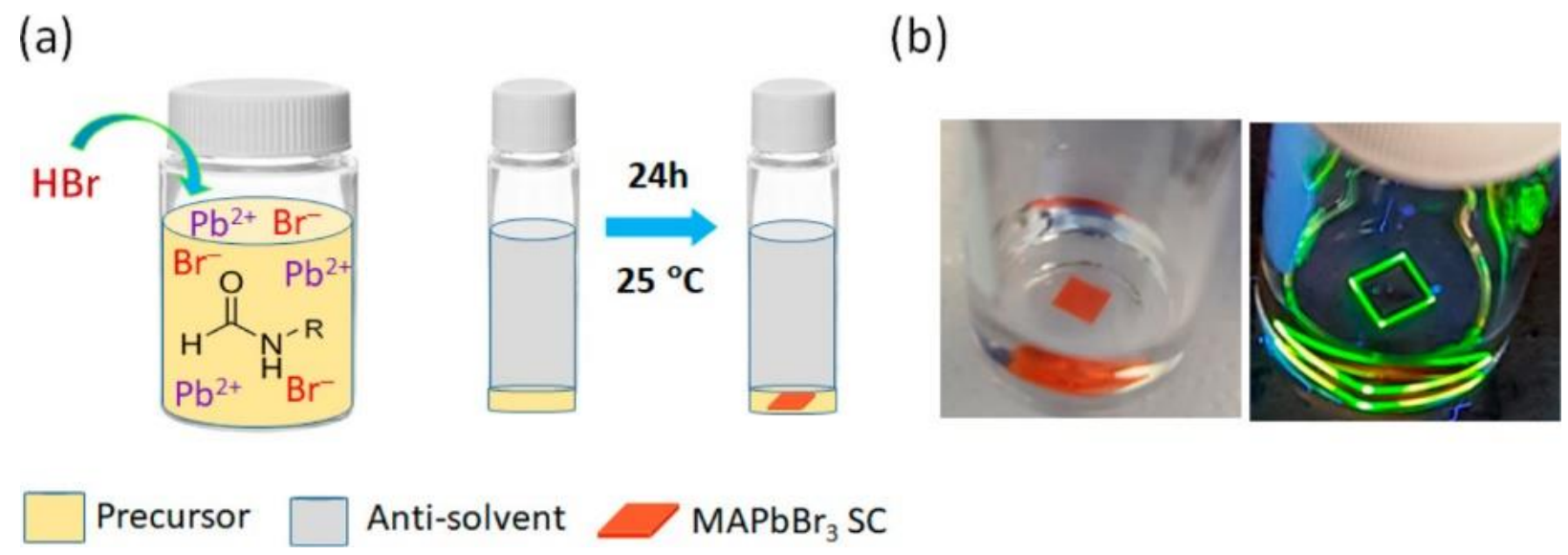

Figure 1. (a) Schematic representation of $MAPbBr_3$ perovskite single crystal growth and (b) photographs of the $MAPbBr_3$ perovskite single crystal under normal light and under a UV lamp.

The crystals showed strong emission at all four edges under UV excitation (Figure 1b). The intense emission at the edges was not evident at the initial growth stages; however, it was clear at the later stages, when the crystal grew larger (Figure S5). We checked several hypotheses on the origin of this strong emission. Hence, we conducted lasing measurements to explore the possibilities of the crystal behaving as a gain media. However, we did not observe any nonlinear PL enhancement (e.g., amplified spontaneous emission) with increasing pump power from 10 to 200.5 μJ $cm^{-2}$ (Figure S6).

Scanning electron microscope (SEM) images showed smooth and uniform morphology with no grain boundaries, even at the edges of the crystals (Figure S7); hence, we excluded the formation of nanograins as the origin of the enhanced edges emission. Energy-dispersive X-ray spectroscopy (EDX) analysis indicated that the Br/Pb atomic ratio at the edges and central regions are 2.93 ± 0.02 and 2.74 ± 0.04, respectively (Figure S8). The higher Br/Pb ratio at the edges suggests a reduced number of structural defects such as bromine vacancies. (61,62)

The crystals exhibited a sharp band edge cutoff (i.e., a reduced Urbach tail, Figure S9) indicating suppressed density of in-gap defect states. (63) The sharp absorption edge at 570 nm corresponds to a bandgap of 2.18 eV, as estimated from the Tauc plot, in good agreement with previous reports. (25,27) Nevertheless, a previous report indicated that the bandgap of such a crystal is usually underestimated. (29)

Steady-state PL measurements were also performed at the strong-emissive and the less-emissive regions. Figure 2a shows the PL spectra of the $MAPbBr_3$ single crystal collected by focusing the 405 nm laser spot (spot diameter of 1.3 mm) at the edge and central area, demonstrating that both regions are emissive; yet, the PL differs in terms of intensity and spectral position. We observed an intense PL emission at 552 nm and a weak PL peak at 538 nm for the edges and central regions, respectively. Based on fluorescence confocal images, the width of the intense emissive edges was found to be 100–120 μm (Figure 2b). In order to gain insight into the PL emission properties in a more localized scale, we carried out μ-PL measurements (Figure 2c). μ-PL emission at 540 nm was observed, and a reproducible spectrum was obtained independently of the point chosen in the central areas of the crystals. On the other hand, PL emission in the range 545–553 nm was observed at the sample edges and side-faces. This trend was noticed by measuring four crystals from different synthesis batches. The red-shifted emission peaks observed in both macro- and μ-PL measurements indicate bandgap narrowing at the edges; hence, shallower traps are expected. (48,51)

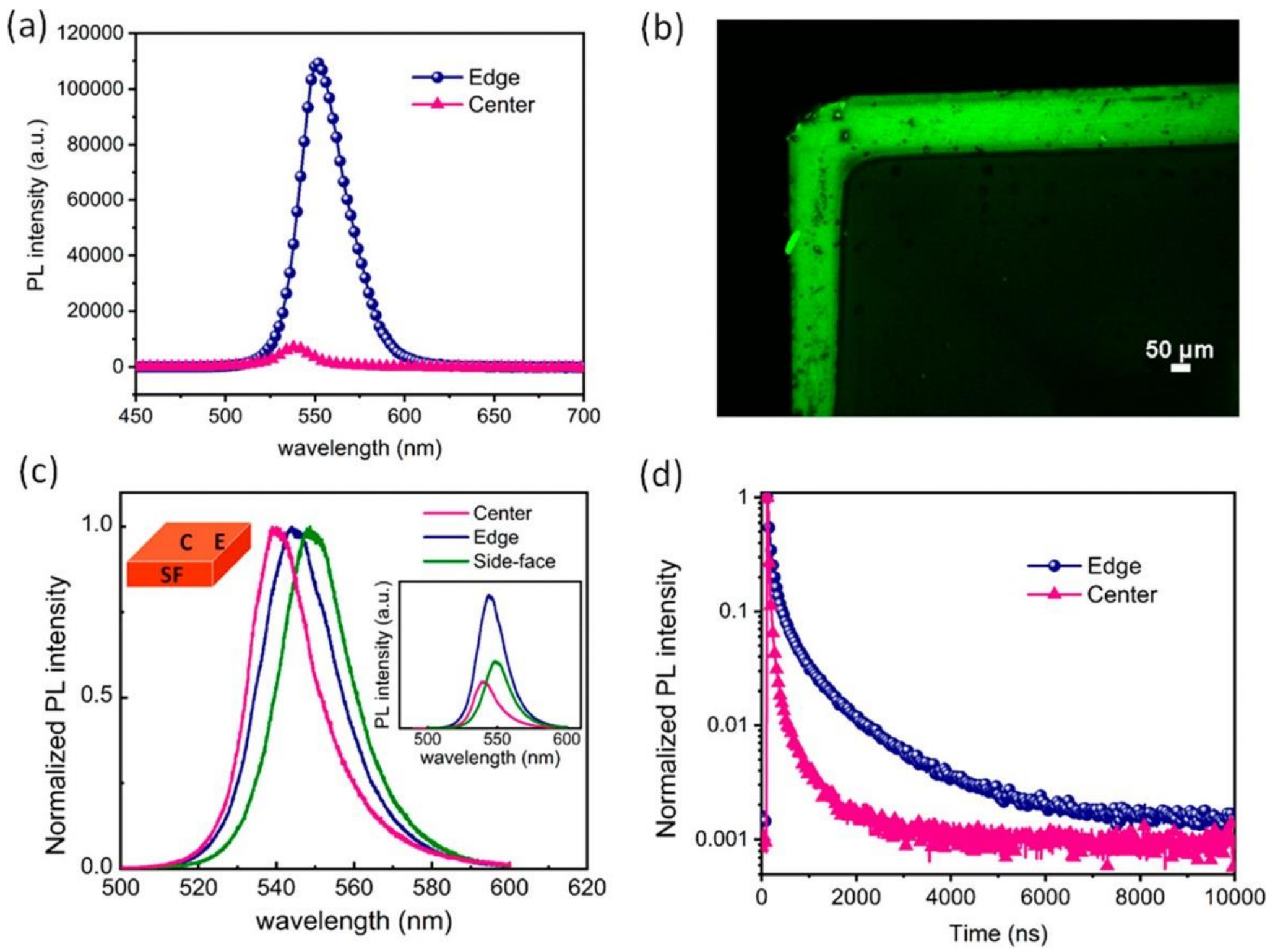

Figure 2. (a) Photoluminescence spectrum of the $MAPbBr_3$ single crystal at the edge and center excited at 405 nm. (b) Fluorescence confocal images of edge emissive $MAPbBr_3$ single crystal. (c) Representative normalized μ-PL emission spectra collected at room temperature under 457 nm laser excitation in different regions of the $MAPbBr_3$ single crystal. Inset: μ-PL spectra without normalization (the sketch in the inset represents the measurement position). (d) Normalized PL decay traces of the $MAPbBr_3$ single crystal at the edge and center, excited at 405 nm and collected at the PL emission peak of 552 and 538 nm with 1 nm bandwidth, respectively.

Time resolved PL (TRPL) spectroscopy was used to investigate the recombination dynamics of photoexcited species by focusing the laser beam onto the different regions (see Figure 2d). The PL decay trace was significantly different at the edge and central regions. Short and long components of 83 ± 21 ns and 1409 ± 119 ns were obtained for the strong-emissive edges, and 11 ± 1 ns and 139 ± 16 ns for the less emissive central parts. Moreover, the contribution of the short component that is due to the nonradiative decay (64) was reduced moving from edges to central areas. These results are consistent with the expected shallower trap states due to the narrower bandgap and also suggest reduced trap state densities at the edges of the crystals compared to their central regions.

We have also cleaved the $MAPbBr_3$ crystal and measured both steady state PL and TRPL at the newly formed edge. As can be seen in the inset in Figure S10, the newly formed edge did not show enhanced emission under a UV lamp. The PL peak position is identical to that measured at the center of the crystals (539 nm). Similarly, PL lifetimes recorded at the newly formed edge (32 ± 2 ns and 76 ± 8 ns) are in the same range as the values measured at the center of the crystals. Therefore, any edge-related structural reorganization can be dismissed as a possible cause of the PL emission.

In order to investigate the nature of the $MA^+$ cation and the $PbBr_3^-$ interactions in $MAPbBr_3$ perovskite crystals, we carried out Raman experiments using an excitation wavelength of 785 nm (i.e., lower energy than the bandgap) to avoid the fluorescence background. As is shown in Figure 3a-b, no shift in Raman modes and no changes in the full-width half-maximum (fwhm) were observed. However, we found that the relative intensity of the different Raman modes collected at the center, edge and side-face of the crystal vary (Figure 3c). The $I_1$(157 $cm^{-1}$)/$I_2$(323 $cm^{-1}$) ratios increased, while the $I_3$(920 $cm^{-1}$)/$I_4$(969 $cm^{-1}$) ratio decreased when moving from the center to the edge and side-face, indicating a different crystalline long-range order within the crystal. (65) All measurements performed so far, including EDX elemental analysis, μ-PL and Raman spectra, indicate a possible variation in the crystal lattice between the edges and central regions of the crystals.

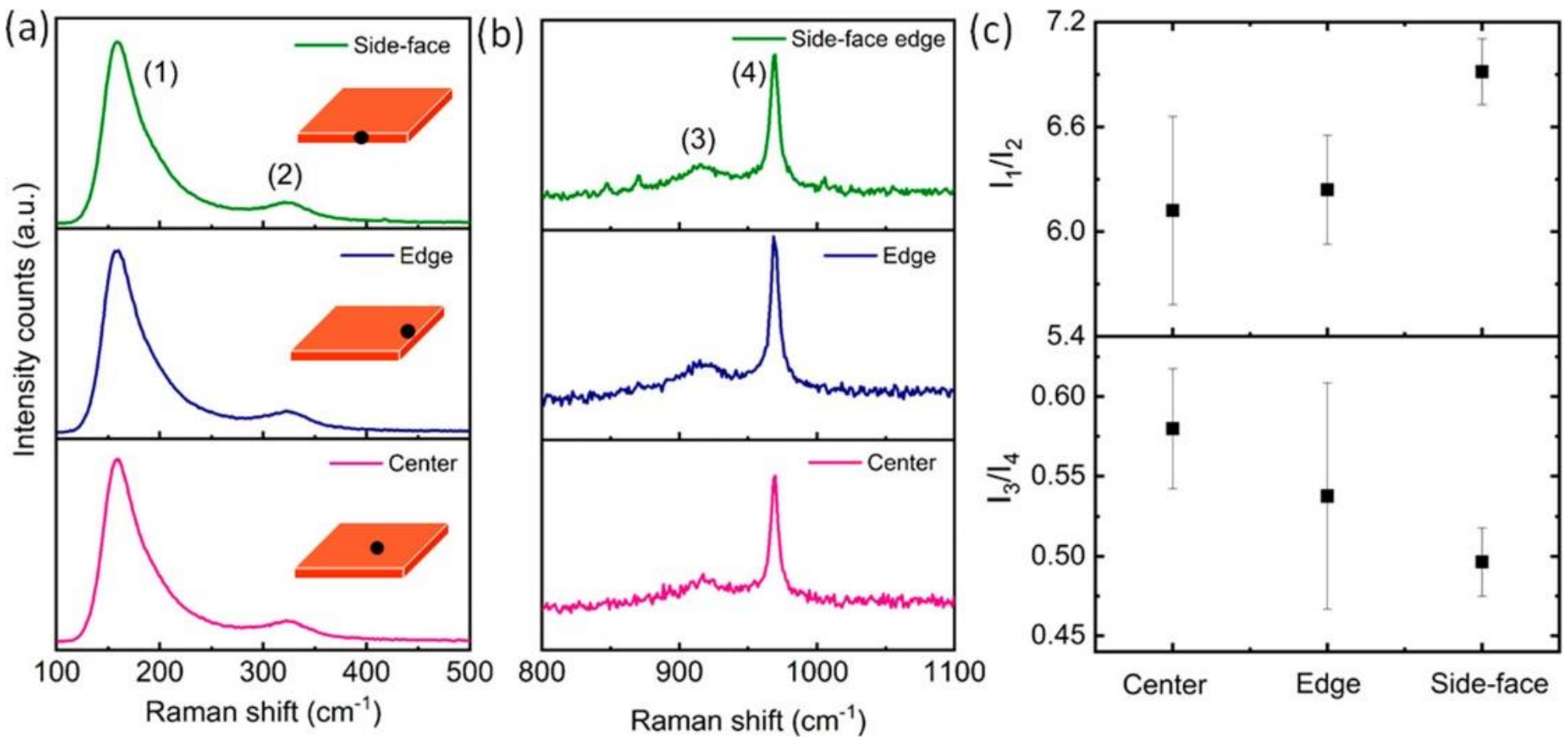

Figure 3. (a–b) Raman spectra of the sample collected at 785 nm excitation wavelength at different regions indicating the four modes observed (157 $cm^{-1}$, related to the vibrations of $PbBr_3^-$ cage and/or coupled motions of $CH_3–NH_3^+$; 323 $cm^{-1}$, C–N torsion ($CH_3–NH_3^+$ rotation); 920 $cm^{-1}$, $CH_3–NH_3^+$ rocking; and 969 $cm^{-1}$, C–N stretching) (66−69) (the sketches in the insets represent the measurement position). (c) Corresponding relative intensities of the different Raman modes collected at the center, edge and side-face of the crystal.

In order to study the structural changes, we deposited a 60 nm thick Au layer on top of the crystal and used the Au (111) X-ray diffraction peak at 38.18° as a reference (Figure S11) to eliminate any peak shift due to height variation (sample z-displacement). The X-ray diffraction (XRD) patterns at the edges of the as-grown perovskite single crystals show diffraction peaks at 15.05°, 30.21°, and 45.94°, which are shifted toward higher angles (around 0.05°) compared to the diffraction peaks from the central regions (15.0°, 30.13°, and 45.89°) as shown in Figure 4.

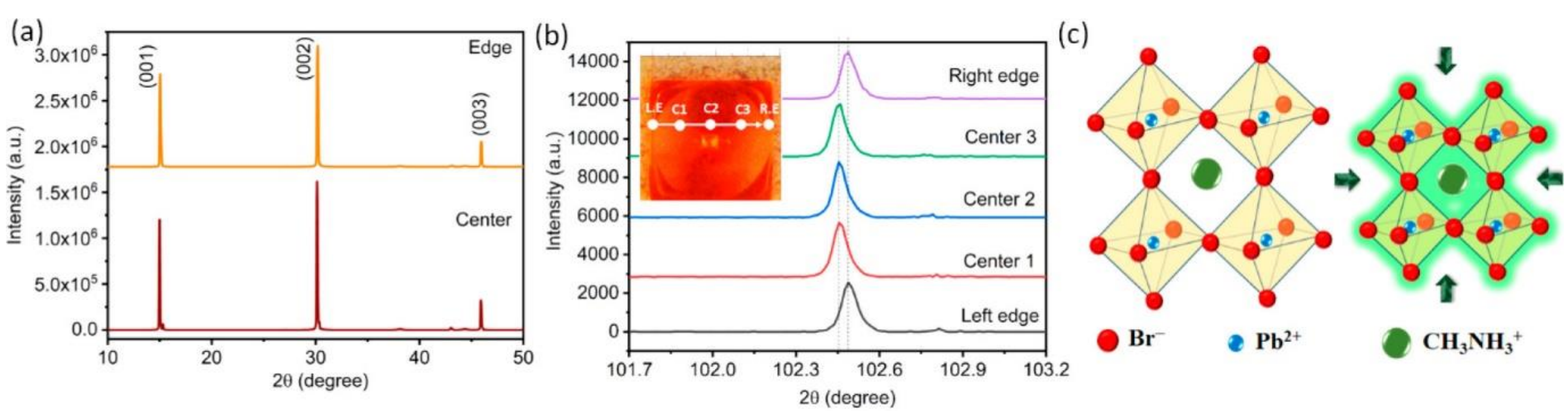

Figure 4. (a) XRD patterns of as-grown $MAPbBr_3$ perovskite single crystal at the edge and center. (b) Line scan analysis through μ-XRD at (006) peak from left edge (L.E.) to right edge (R.E.) of the millimeter size single crystal (Inset: photograph of $MAPbBr_3$ single crystal with μ-XRD line scan measurement position). (c) Schematic representation of perovskite structure without (left) and with (right) lattice compression.

We also performed μ-XRD line scan analysis on our sample with a spatial resolution of about 100 μm to further confirm the structural difference at the edges and the central regions. The μ-XRD patterns were collected for higher angle peaks corresponding to the (004), (005), and (006) planes, as is shown in Figures 4b and S12. Note that choosing higher angle diffraction peaks allows minimization of the X-ray spot on the sample. We observed a systematic shift to higher angles in all three peak positions in the diffraction patterns collected from the edges, compared to that of the central parts of the crystals. This result confirms lattice compression and reduction in the unit cell (Figure 4c) of about 0.27% at the edges with respect to central parts of the crystals. This lattice compression is directly associated with bandgap narrowing, shallower traps, and reduced nonradiative recombination pathways in halide perovskites, hence the observed intense and red-shifted emission, and the long PL lifetime at the edges. (48,50,51,56,57)

The perovskite precursors and their relative ratios are known to have remarkable effect on their nucleation and growth dynamics. (70,71) As mentioned before, in our process, there is a gradual release of the MA ions which would keep changing the concentration and the relative ratio between the perovskite precursors. These changes could be the reason behind the compressed edges.

To understand how this irreversible compression observed at the edges affects the optoelectronic properties, we fabricated planar photodetectors using our AA-SAC grown crystals. The device structure is shown in Figure 5a. Ti electrodes were deposited using e-beam evaporation at the edge and central regions. Figure 5b and c show the current–voltage (*I*–*V*) curves of a typical photodetector fabricated employing the edge and center of the crystal, respectively, under varying illumination intensities (from $0.15 \times 10^{-3}$ to 5 mW $cm^{-2}$) using a 473 nm laser and a different applied bias (−5 to 5 V). The photodetector devices exhibited dark current as low as $8.89 \times 10^{-10}$ A and $17.11 \times 10^{-10}$ A at the edges and central areas, respectively, at 5 V bias (Figure S13). As the device with lower dark current is expected to have fewer defects, (26,72) we conclude that there are fewer defects at the edges than at the central regions. The detectivity ($D^*$), responsivity ($R$), and linear dynamic range (LDR) of the photodetectors were calculated by measuring the photocurrent under different illumination intensities (Note S1, Supporting Information). The $D^*$ is calculated as a function of the illumination intensity and ranged from $0.15 \times 10^{-3}$ to 5 mW $cm^{-2}$ at a laser wavelength of 473 nm with a fixed 5 V bias (Figure 5d). A high $D^*$ of $(2.73 \pm 0.28) \times 10^{13}$ cm $Hz^{1/2}$ $W^{-1}$ (Jones) was recorded at the edges, which is more than five times higher than the value ($(5.17 \pm 0.22) \times 10^{12}$ cm $Hz^{1/2}$ $W^{-1}$ (Jones)) recorded at the central areas under the incident power $0.15 \times 10^{-3}$ mW $cm^{-2}$. Again, we ascribed the higher detectivities at the edges to a lower density of defects, as indicated by the lower dark current. The maximum values of $R$ were 80.8 ± 3.8 A $W^{-1}$ and 51.5 ± 6.3 A $W^{-1}$ at the edges and the central regions, respectively (Figure 5e). The photocurrent response was also measured for the $MAPbBr_3$ single crystal devices at the edges and central parts, as a function of wavelength (Figure S14). It is worth noting that the edges showed an extended spectral response of about 5 nm, which confirms the bandgap narrowing at the edges due to the lattice compression. The enhanced performance at the edges holds in terms of transient measurements and LDR, as summarized in Figure S15 and the accompanied discussion. Using Au electrodes, instead of Ti, resulted in similar behavior; however, the device with the Ti electrode exhibited lower dark current (Figures S16, S17 and accompanied discussion). Finally, using space charge limited current (SCLC) measurements (Figure 5f and Note S2), we confirmed the reduced trap state densities at the edges ($(1.87 \pm 0.43) \times 10^{11}$ cm$^{-3}$) compared to the central areas ($(6.89 \pm 0.96) \times 10^{11}$ cm$^{-3}$), which is in agreement with the enhanced PL and longer carrier lifetime as observed from TRPL spectra.

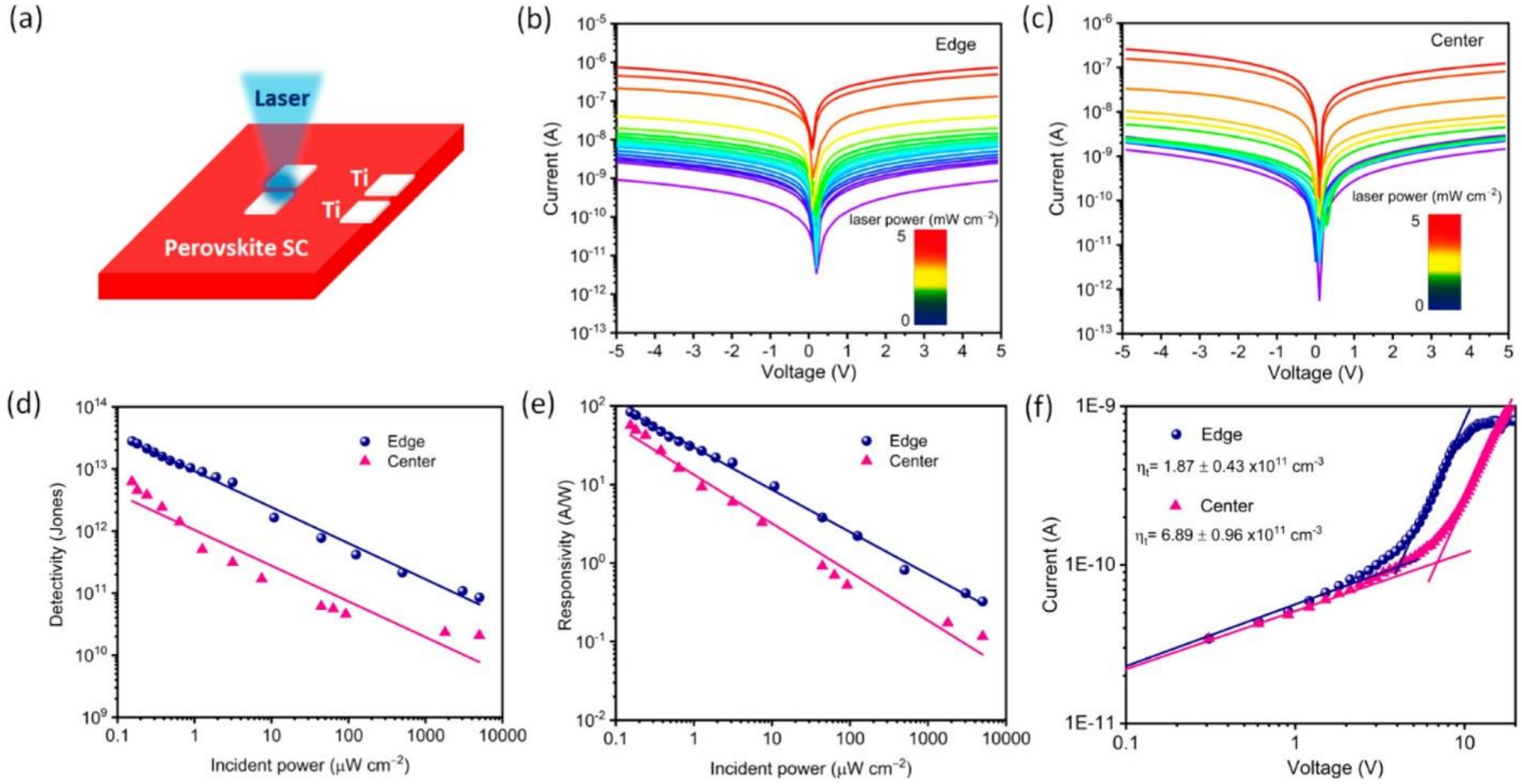

Figure 5. (a) Schematic of perovskite single crystal photodetector device. *I–V* curves of perovskite photodetectors fabricated at (b) the edge and (c) the center under varying illumination intensities ranging from $0.15 \times 10^{-3}$ to 5 mW cm$^{-2}$ using a 473 nm laser and bias (−5 to 5 V). Power-dependent (d) detectivity and (e) responsivity of the $MAPbBr_3$ single-crystal photodetector at 5 V using Ti electrode. (f) The dark *I–V* curve of $MAPbBr_3$ single crystal based on a hole-only device.

In summary, we have prepared high-quality $MAPbBr_3$ perovskite single crystals using a room temperature AA-SAC method. μ-XRD, μ-PL, and μ-Raman analyses confirmed that the $MAPbBr_3$ edges of the crystals are structurally compressed with respect to the central regions. Consequently, the edges of the crystals exhibited slightly red-shifted and much enhanced emission with long PL lifetimes. Furthermore, we recorded excellent detectivities and high photoresponsivities at the edges. Our results suggest that by controlling the crystallization process we can reduce both structural defects and nonradiative pathways, thus improving the radiative recombination of perovskite based devices.

**Acknowledgments**

We thank Dr. F. De Angelis (Plasmon Nanotechnologies Group - IIT) for access to the Raman equipment for the measurements and IIT Clean Room Facility for access to the equipment necessary for device fabrication. We thank Dr. Roman Krahne (Optoelectronics group - IIT) for access to probe station from Janis Research for photodetector measurement. We thank Simone Lauciello (Electron Microscopy Facility - IIT) for collecting the SEM images.